\documentclass[10pt, conference, letterpaper]{IEEEtran}
\IEEEoverridecommandlockouts
\usepackage[hyphens]{url}
\usepackage{cite}
\usepackage{amsmath,amssymb,amsfonts}
\usepackage{algorithmic}
\usepackage{graphicx}
\usepackage{textcomp}
\usepackage{xcolor}
\def\BibTeX{{\rm B\kern-.05em{\sc i\kern-.025em b}\kern-.08em
    T\kern-.1667em\lower.7ex\hbox{E}\kern-.125emX}}

\usepackage[pscoord]{eso-pic}
\usepackage{xparse}
\usepackage{tikz}
\usepackage{annotate-equations}
\usepackage{mathtools}
\usepackage{array}
\usepackage{xcolor}
\definecolor{LightGray}{gray}{0.9}
\usepackage{listings}
\usepackage{graphicx}
\usepackage{booktabs}
\usepackage{acronym}
\usepackage[export]{adjustbox}
\usepackage{balance}
\usepackage{pifont}
\usepackage{framed}
\usepackage{soul}
\usepackage{csquotes}
\usepackage{multirow}
\usepackage{array}
\usepackage{algorithm2e}
\usepackage{algorithmic}
\usepackage{graphicx}
\usepackage{booktabs}
\usepackage{mathtools}
\usepackage{acronym}
\usepackage[draft]{minted}
\SetKw{Continue}{continue}
\newcolumntype{L}[1]{>{\raggedright\let\newline\\\arraybackslash\hspace{0pt}}m{#1}}
\newcolumntype{C}[1]{>{\centering\let\newline\\\arraybackslash\hspace{0pt}}m{#1}}
\newcolumntype{R}[1]{>{\raggedleft\let\newline\\\arraybackslash\hspace{0pt}}m{#1}}
\MakeOuterQuote{"}
\usepackage[caption=false,font=footnotesize,labelformat=simple,listofformat=subsimple]{subfig} % caption=false is needed for IEEE template

\acrodef{IP}[IP]{Intellectual Property block}
\acrodef{SoC}[SoC]{System-on-Chip}
\acrodef{PC}[PC]{Patching Controllability}
\acrodef{PO}[PO]{Patching Observability}
\acrodef{RTL}[RTL]{Register Transfer Level}
\acrodef{FPGA}[FPGA]{Field Programmable Gate Array}
\acrodef{eFPGA}[eFPGA]{embedded Field Programmable Gate Array}
\acrodef{PC}[PC]{Patching Controllability}
\acrodef{alg}[PIDP]{Patch Infrastructure Design Problem}
\acrodef{ALM}[ALM]{adaptive logic module}

\DeclarePairedDelimiter\floor{\lfloor}{\rfloor}
\lstdefinestyle{verilog-style}
{
    language=Verilog,
    basicstyle=\small\ttfamily,
    keywordstyle=\color{vblue},
    identifierstyle=\color{black},
    commentstyle=\color{vgreen},
    numbers=left,
    numberstyle=\tiny\color{black},
    numbersep=10pt,
    tabsize=8,
    moredelim=*[s][\colorIndex]{[}{]},
    literate=*{:}{:}1
}
\usepackage{diagbox}
\newlength{\mintednumbersep}
\AtBeginDocument{%
  \sbox0{\tiny00}%
  \setlength\mintednumbersep{\parindent}%
  \addtolength\mintednumbersep{-\wd0}%
}
\usepackage{tikz}

\author{%
    \IEEEauthorblockN{Wei-Kai Liu\IEEEauthorrefmark{1}, Benjamin Tan\IEEEauthorrefmark{2}, Jason M. Fung\IEEEauthorrefmark{3}, and Krishnendu Chakrabarty\IEEEauthorrefmark{4}\thanks{This work was supported in part by the Semiconductor Research Corporation (SRC) under Task 3065.001 and Task 3198.001. This work does not in any way constitute an Intel endorsement of a product or supplier.}}
    \IEEEauthorblockA{%
    \IEEEauthorrefmark{1}%Department of Electrical and Computer Engineering, 
    Duke University,
    \IEEEauthorrefmark{2}%Department of Electrical and Software Engineering, 
    University of Calgary,
    \IEEEauthorrefmark{3}Intel Corporation,
    \IEEEauthorrefmark{4}%School of Electrical, Computer and Energy Engineering, 
    Arizona State University \\
    weikai.liu@duke.edu, benjamin.tan1@ucalgary.ca, jason.m.fung@intel.com, krishnendu.chakrabarty@asu.edu
    }
}

\usepackage{geometry}
\begin{document}

\bstctlcite{IEEEexample:BSTcontrol}

\title{Theoretical Patchability Quantification for IP-Level Hardware Patching Designs}
%in SoCs}

\maketitle

\begin{abstract}
As the complexity of System-on-Chip (SoC) designs continues to increase, ensuring thorough verification becomes a significant challenge for system integrators.
The complexity of verification can result in undetected bugs. 
Unlike software or firmware bugs, hardware bugs are hard to fix after deployment and they require additional logic, i.e., patching logic integrated with the design in advance in order to patch.
However, the absence of a standardized metric for defining "patchability" leaves system integrators relying on their understanding of each IP and security requirements to engineer ad hoc patching designs.
In this paper, we propose a theoretical patchability quantification method to analyze designs at the Register Transfer Level (RTL) with provided patching options. 
Our quantification defines patchability as a combination of observability and controllability so that we can analyze and compare the patchability of IP variations.
This quantification is a systematic approach to estimate each patching architecture’s ability to patch at run-time and complements existing patching works. 
In experiments, we compare several design options of the same patching architecture and discuss their differences in terms of theoretical patchability and how many potential weaknesses can be mitigated.
% \textit{Note: 8 $+$ 1 pages max. for ICCAD. You must enter plain text only, and you may not exceed 250 words.}
\end{abstract}
% \begin{IEEEkeywords}
% Patching, System-on-Chip, Hardware Security
% \end{IEEEkeywords}

% NOTE: REMOVE BEFORE SUBMISSION
% \thispagestyle{plain}
% \pagestyle{plain}

\section{Introduction}
As SoC complexity continues to increase, system integrators face significant challenges in ensuring thorough design verification~\cite{chen2017challenges, trippel2021bomberman}.
It is difficult for system integrators to simulate and test all possible scenarios, allowing some bugs to go undetected until deployment~\cite{charles2020real,ahmad2022don}. 
This challenge is exacerbated by time-to-market pressures, where verification teams must work under tight schedules. %, leading to corner cases being bypassed, and some bugs not being adequately tested. 
As bugs can be found in hardware~\cite{dessouky2019hardfails,delshadtehrani2020phmon}, software~\cite{makhshari2021iot}, or firmware~\cite{ray2019formal}, detecting them can require a mix of verification techniques, such as simulation, emulation, and formal approaches, which further increases the difficulty of detecting all bugs during design.

\begin{figure}
    \centering
    \includegraphics[width=0.62\columnwidth]{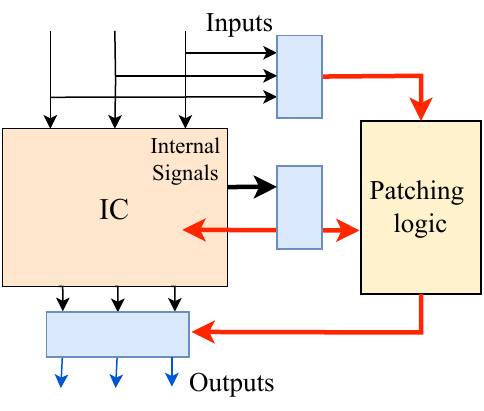}
    % \vspace{-10pt}
    \caption{The general concept of adding patching logic to improve hardware patchability. Red (thick) arrows going toward the IC indicate modifications made by the patching logic, while arrows going toward the patching logic indicate observations made by the patching logic. Black (thin) arrows represent the inputs, internal signals, and outputs of the IC.}
    \label{fig:overview}
    % \vspace{-3mm}
\end{figure}

While updates can often address firmware/software bugs~\cite{kollenda2018exploratory,dharsee2017software}, hardware bugs are typically considered as ``unfixable'' after deployment. 
% The MITRE Corporation has built a class of hardware bugs under the Common Weakness Enumeration (CWE) list~\cite{the_mitre_corporation_mitre_cwe-1194_2021} that require additional logic integrated with the designs to patch.
% To avoid malicious users from utilizing the covert bugs, 
To improve the resilience of systems in the field, approaches have been proposed to insert hardware-supported patching infrastructure to provide in-field patching actions~\cite{sarangi2006phoenix, hicks2015specs, delshadtehrani2020phmon, tan2020toward, nath2018system}.
However, such solutions are bespoke as it is not straightforward to compare different candidate solutions for their patching ability (which we refer to as ``patchability''). 
In this work, we consider a patching solution that involves modifying IPs with added ``patching logic'' such that some signals in the design can be observed and/or modified at run-time by patching hardware, as shown in Fig.~\ref{fig:overview}.
% However, at the design stage, where bugs remain undetected, system integrators would not know how vulnerable each \ac{IP} is, what kind of patching infrastructure they need, and which one is better than the other in terms of the patching ability (patchability).

The desire for more efficient designs limits resources available for patching, thus motivating a trade-off between patchability and resource investment. 
This poses a major challenge for \ac{SoC} designers. 
Without a metric for defining patchability, system integrators must rely on their understanding of each IP and security requirements to engineer ad hoc patching designs~\cite{liu2022hardware}. 
Even if the resources available to invest in the patching architecture are determined, it is unclear which signals or registers require patching logic to achieve higher patchability. 
As a result, patching architectures are often suboptimal in terms of patching potential bugs.

In this work, we propose an approach to quantify the theoretical patchability of a design at \ac{RTL} given several patch logic options.
To the best of our knowledge, this is the first attempt to define patchability for patching hardware as a metric that combines notions of observability and controllability. %, analyzing how much they can be given by the patching design at a specific \ac{IP}.
This quantification complements prior patching work by introducing an estimate of different patching infrastructures that can be used in design trade-offs. % to patch at run-time.
% In experiment, we compare a few design options of the same patching infrastructure and discuss their difference in terms of the theoretical patchability, resource usage, and how many potential weaknesses can be mitigated.
Our contributions are as follows:
\begin{itemize}
    \item We propose a novel definition of patchability for RTL designs with integrated hardware patching logic. 
    % to analyze RTL designs; %define what does patchability mean for hardware patching
    % \item propose a patchability quantification method to analyze designs at RTL
    \item We present the process to determine patchability scores for RTL designs, enabling designers to explore the effect of exposing different internal signals to patching logic. 
    \item Through the evaluation of an IP case study, our patchability scoring analysis demonstrates that our proposed approach achieves a normalized score of 0.9 while utilizing 65\% fewer resources compared to a greedy approach.
\end{itemize}

\section{Background}
% This section introduces definitions that are used throughout this work. 
% First, we begin by elaborating our definition of a ``patch''.
% While our focus is mainly on hardware blocks, the concept can apply to computer-based systems as a whole. 
% Next, we define ``patchability'' at the RTL, where SoC developers design patching infrastructure and insert patching logic. 
% We also describe how a patching action is executed in hardware.
% Since patches are written after bugs are discovered, the patching infrastructure must be reprogrammable so that patches can be updated.

% \subsection{What is a Patch?}
We consider a \textit{patch} to be a sequence of actions introduced after deployment that can mitigate an issue. We classify patches as fixes (design changes to eliminate the buggy operation) and workarounds (design changes that modify the bug's effects, without necessarily fixing the actual cause). 
% For example, consider an IP that occasionally enters an unintended terminal state. A workaround patch can be configuring hardware to reset only the buggy IP; from an external viewpoint, the bug is addressed even if the IP itself can still enter the undesired state. 
Software or firmware-based patches can update/fix bugs in the field, such as updating access control policy or microcode~\cite{kollenda2018exploratory}.
To achieve something similar for, and with hardware, hardware-based patches need hardware that is pre-designed to enable reconfiguration, i.e., inserted patching logic. 
Typically, a hardware patch comprises two parts: observation logic which identifies the buggy scenario and determines when to launch the corrective actions, and correction logic, which takes action to mitigate issues once notified~\cite{tan2020toward,nath2018system,liu2022hardware}. 
A \textit{hardware-based} patch requires an understanding of hardware operation and the ability to observe and control (modify) hardware signals~\cite{liu2022hardware}. 

\begin{figure}[t]
    \centering
    \includegraphics[width=0.65\columnwidth]{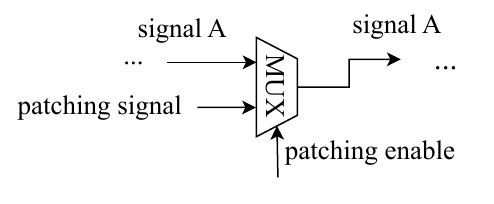}
    \caption{Adding a patching control cell involves replacing an internal wire with a MUX and a patching signal, enabling external patching logic to modify the signal behavior.}
    %The top figure shows the original connectivity of signal A, while the bottom figure illustrates the connectivity after inserting a patching control cell where a MUX and two patching relevant signals are required.
    \label{fig:patchingdef}
    % \vspace{-1em}
\end{figure}

To implement hardware patches, we consider a \textit{patching control cell} as part of the patching mechanism. 
When selecting a signal to be directly patchable (modifiable), we substitute the original signal wire with a patching control cell (Fig.~\ref{fig:patchingdef}). % is used to substitute the original signal wire.
% A patching control cell consists of an enable signal that activates the patch, switching the original signal to the patching signal (Fig.~\ref{fig:patchingdef}). 
To complete the patching process, the patching enable is pulled high and the patching signal is driven with the appropriate value to modify the original behavior of the signal. 
Patching control cells can replace inputs, outputs, internal registers, or even constant signals, to augment design elements that are already reprogrammable (e.g., control registers). 
As mentioned, a patch typically comprises an observation and correction component -- these determine the \textit{patchability} of a patching design. 
Patchability, therefore, has two key features: observability and controllability. %, which correspond to the two components of a patch.
The concepts of observability and controllability have been proposed in previous work~\cite{goldstein1980scoap} and are widely used to determine a design's testability. 
For clarity, we will refer to our concepts as \ac{PO} and \ac{PC}, respectively.

% The main difference between our \ac{PO}/\ac{PC} and observability/controllability is that while the latter is designed to model observing the outputs and controlling the inputs of a design during testing, our approach focuses on observing and controlling any signals within the design for patching purposes. 
% It is important to note that f
% For third-party IPs where we might not have direct internal access, we can only observe and control signals at the interface level.
% Additionally, as patching architectures need to be designed at the RT-level where system integrators have more insight into IPs (e.g., know which signal is more important), \ac{PO} and \ac{PC} analysis must be done at this level while the traditional quantification is used on gate-level netlists.
% \subsection{}

% \subsection{Patching Controllability and Patching Observability}
Patching Controllability (PC) refers to the extent to which a signal can be controlled by the patching logic.
% For an $n$-bit signal, PC can range from $0$ to $n$, where $n$ represents the maximum number of bits that can be controlled. 
For a 1-bit signal, $i$, PC is the probability that the bit can be controlled, where 0 means the bit cannot be set to a chosen value, while 1 means the bit can be set to either 1 or 0, i.e., $0 \leq PC_i \leq 1$. 
For an $n$-bit signal $\vec{v}$ (e.g., a bus or bit-vector), we define its PC, $PC_{\vec{v}}$ to be the product of the probability of controlling each bit and the number of bits in the signal, i.e., $0 \leq PC_{\vec{v}} \leq n$. 
For simplicity, we assume that all bits in an $n$-bit signal have the same bit-level PC. 
% To calculate the PC score for an $n$-bit signal (e.g., a bus or bit-vector), $\vec{v}$, we take the expectation of the number of bits that can be controlled, which is equal to the product of the probability of controlling each bit and $n$.
% Here, we assume that each bit of a signal has equal probability of being controlled.
% When a signal is being controlled by the patching logic, it is assumed that this signal can be changed to any desired value, resulting in a PC score of $n$ for an $n$-bit signal connected directly to the patching logic.  
% Therefore, for an $n$-bit signal, the PC score ($PC_{score}$) is defined as:
% \begin{equation}
% PC_{score} = \sum_{i=1}^{n} p_i,    
% \end{equation}
% where $p_i$ is the probability of controlling the i-th bit.
% The maximum PC score for an $n$-bit signal is $n$, which represents "fully controllable" and occurs when all bits are controllable (i.e., $p_i = 1$ for all $i$).
% The minimum PC score for an $n$-bit signal is $0$, which represents "uncontrollable" and occurs when none of the bits are controllable (i.e., $p_i = 0$ for all $i$).
Patching Observability (PO), which refers to the extent to which a signal can be observed by the patching logic, has the same formulation as \ac{PC}.

For this work, we assume that all patching signals are generated using reprogrammable logic (e.g., Lookup Tables (LUTs) or some other structure such as a patching block~\cite{liu2022hardware}). 
This means that when an $n$-bit signal is replaced by a patching control cell, the \ac{PC} score is considered to be $n$, indicating that the patch logic can change all $n$-bits to a chosen value. 
This is defined as ``fully controllable.''

\section{Patchability Formulation \label{sec:formulation}}
This section outlines how theoretical patchability can be quantified at the RTL. 
As the quantification of PC and PO are mathematically equivalent, we will only focus on the calculation of PC in this paper.
\begingroup
\renewcommand{\arraystretch}{1.5} % Default value: 1
\begin{table}[hbtp!]
\caption{PC score calculation for basic operators}
    \centering
     \begin{tabular}{|c|c|}
    \hline
        Operator & PC score \\ \hline
        ASSIGN/NOT & $S_A$\\ 
        SHIFT & $S_A\times \frac{n-1}{n}$\\ 
        OR/NOR  &  $\frac{S_A+S_B}{2}$\\ 
        AND/NAND & $ \frac{S_A+S_B}{2}$\\ 
        XOR/XNOR & $ \frac{S_A+S_B}{2}$\\ 
        CONCATENATE & $S_A+S_B$\\ \hline
        % DECONCATENATE & $ $\\ \hline
    % \bottomrule
    \end{tabular}
    \label{tab:operator}
    % \vspace{-2em}
\end{table}
\endgroup

\subsection{Basic Operators}
To analyze patchability at the RTL, we explain how the PC score is calculated given the logic of a design. % change during operations.
% Here we show the remaining PC score for some common operations in RTL in Table~\ref{tab:operator}. 
We show the PC score for some common operations in RTL in Table~\ref{tab:operator}. 
For the two-input operators, $S_A$ and $S_B$ represent the PC scores of the input signals.
For the single-input operator, we use $S_A$ to refer to the score of the input signal.
For the nets that hold constant values, their PC scores are defined as 0 because we cannot control these values.
% Table~\ref{tab:operator} shows the remaining PC score for $n$-bit signals of a selection of common operators in RTL. 
% The probability that each bit of input1/input2 can be controlled is $\frac{in_1}{n}$/$\frac{in_2}{n}$.
% \begin{equation*}
%     \begin{aligned}
%         & 10111 \;\; \text{(input1)}\\
%         +& 01101 \;\; \text{(input2)}\\
%         -&---\\
%         &11111 \;\; \text{(out)}
%     \end{aligned} 
% \end{equation*}
We demonstrate their derivation. %se are derived.

\textit{\textbf{OR operation} (with two inputs A and B and one output out)}\\
% \textit{OR operation (Consider an OR gate with two inputs A and B)}\\
% \textit{OR operation (Fig. \ref{fig:or}):} 
% \begin{figure}[htbp!]
%     \centering
%     \includegraphics[width=0.4\columnwidth]{figure/OR.png}
%     \caption{OR gate}
%     \label{fig:or}
% \end{figure}
The probability that each input A and B (both 1-bit) is controllable is $S_A$, $S_B$, respectively.
% Therefore, the probability that two controllable bits are aligned is 
Therefore, the probability that the two inputs are both controllable (assuming the signals are independently controllable) is:
\begin{equation}
    % \frac{in_1}{n}\times \frac{in_2}{n}.
    S_A\times S_B.
    \label{eq:bothcontrol}
\end{equation}
The probability that only one input is controllable is
\begin{equation}
    % \frac{in_1}{n}\times (1-\frac{in_2}{n})\;+\; (1-\frac{in_1}{n})\times \frac{in_2}{n}
    S_A\times (1-S_B) + (1-S_A)\times S_B,
\end{equation}
where we assume that each bit has an equal probability of being 1 and 0, e.g., $\frac{1}{2}$ for both.
Thus, the probability that one bit is controllable while the other is not controllable but is the non-controlling value (e.g., 0 for \textit{OR} operation) is:
\begin{equation}
    % \frac{1}{2}\times (\frac{in_1}{n}\times (1-\frac{in_2}{n})\;+\; (1-\frac{in_1}{n})\times \frac{in_2}{n})
    \frac{1}{2}\times (S_A\times (1-S_B) + (1-S_A)\times S_B)
    \label{eq:onecontrol}
\end{equation}
As a result, the probability that $out$ can be controllable is the summation of Equation~(\ref{eq:bothcontrol}) and Equation~(\ref{eq:onecontrol}), so the resulting PC score of $out$ is
\begin{equation}
\begin{aligned}
    &S_A\times S_B+\frac{1}{2}\times (S_A\times (1-S_B) + (1-S_A)\times S_B)\\
    &=\frac{S_A+S_B}{2}
    % &n\times (\frac{in_1}{n}\times \frac{in_2}{n}+\frac{1}{2}\times (\frac{in_1}{n}\times (1-\frac{in_2}{n})\;+\; (1-\frac{in_1}{n})\times \frac{in_2}{n}))\\
    % &= \frac{in_1+in_2}{2}
    \label{eq:remaining}
\end{aligned}
\end{equation}
\textbf{AND} operation can be derived in a similar way (1 is the non-controlling value).
The \textbf{XOR} operation does not have a controlling value. However, if one of the inputs is set to 0, the output depends on the other input. As a result, we can still obtain the same results as for \textbf{OR} and \textbf{AND}.
Since we can identify the corresponding input values by observing the outputs of \textbf{ASSIGN} and \textbf{NOT}, the PC score remains the same after these two operations.
The remaining PC score of other operations can be derived by using a combination of these. % operations.

% \wk{TODO: Plus vs. OR: a = b+c ..?\\}

\subsection{Conditional Assignment}
Next, we discuss how to determine the PC score of conditional assignments. 
These examples are all 1-bit signals, and we will show more complex scenarios later.
For instance, a statement where signal $A$ is conditioned on signal $B$ is shown below: 
\begin{equation*} 
{A} <= (B)\;? \;C:\;D
\end{equation*}
To derive the formulation of conditional assignments, we first consider a few scenarios: \\
% \wk{indent, smaller?}
If $B$, $C$, and $D$ are fully controllable, we can fully control $A$. \\
If only $B$ \& $C$ are fully controllable, we can fully control $A$. \\
If only $B$ is fully controllable, we cannot fully control $A$. \\
If only $C$ \& $D$ are fully controllable, we can fully control $A$. \\
If only $C$ or $D$ is fully controllable, we cannot fully control $A$. \\

% \begin{enumerate}[1.]
%     \item If $B$, $C$, and $D$ are fully controllable, we can fully control $A$.
%     \item If only $B$, $C$ are fully controllable, we can fully control $A$.
%     \item If only $B$ is fully controllable, we cannot fully control $A$.
%     \item If only $C$ and $D$ are fully controllable, we can fully control $A$.
%     \item If only $C$ or $D$ is fully controllable, we cannot fully control $A$.
%     % \item if $C$ and $D$ are 0 and 1 and $B$ is fully controllable, we can fully control $A$.
%     % \item if one of $C$ and $D$ is 0/1 while the other is not a fixed value, and only $B$ is fully controllable, we cannot fully control $A$.
% \end{enumerate}

% \wk{Even though we can represent conditional assignments by AND and OR gates ($A=BC+!BD$) and derive the corresponding PC score, this is not accurate as it is impossible that both $B$ and $!B$ are non-controlling bits simultaneously. Thus we need another formulation.}

We use $S_A$, $S_B$, $S_C$, and $S_D$ to represent the PC score for signals A, B, C, and D, respectively.
To satisfy all the scenarios, we propose the following formulation for the conditional operator:
\begin{equation}
\begin{aligned}
    S_A &= S_B\times max\{S_C,S_D\}+(1-S_B)\times\frac{1}{2}\times (S_C + S_D)\\
    % & = S_B\times max\{S_C,S_D\}+\frac{(1-S_B)}{2}\times max\{S_C,S_D\}+\\
    % &\frac{(1-S_B)}{2}\times min\{S_C,S_D\}\\
    % & =\frac{1+S_B}{2}\times max\{S_C,S_D\}+\frac{1-S_B}{2}\times min\{S_C,S_D\}\\
    % &= \frac{S_C+S_D}{2}+\frac{S_B}{2}\times abs(S_C-S_D)
\end{aligned}
\label{eq:conditional}
\end{equation}
The key idea is that, given the probability $S_B$ that we can control the value of $B$ and thereby choose whether signal $C$ or $D$ is assigned to $A$, we will always choose the signal that has the higher PC score (i.e., $max\{S_C,S_D\}$).
When we cannot control which signal to choose (with the probability $1-S_B$), $B$ has an equal probability to be 0~or~1 to select signals $C$ and $D$.

If there are any constants within a conditional assignment, the PC score in Equation~(\ref{eq:conditional}) needs to be modified.
For instance, in the assignment below:
\begin{equation}
\begin{aligned}
{A} <=& (B)\;? \;0\;(C):\;1\;(D),\\
\end{aligned}
\label{eq:constantcase}
\end{equation}
signals $C$ and $D$ are constants that are defined to have a PC score of $0$. 
However, note that A is fully controllable if $B$ is also fully controllable.
Therefore, to precisely evaluate the PC score in conditional assignments when the select signal is fully controllable ($S_B = 1$), the scores for the constants are changed to:
\begin{equation}
    S_C = S_D = \floor{log_2X},
    \label{eq:casestatement}
\end{equation}
where $X$ is the number of distinct constants in the conditional block.
Equation~(\ref{eq:casestatement}) quantifies the minimum number of bits that are affected by the constants.
In Equation~(\ref{eq:constantcase}), two different values are present, sufficient to contribute a one-bit change of $A$.
Thus, $S_C$ and $S_D$ are 1, and therefore $S_A = S_B$.
This modification is also needed to compute the PC score in case statements.
For example, consider
\begin{equation*}
\begin{aligned}
case(B)&:\\
00&: A<= 00\; // (C)\\
01&: A<= 01\; // (D)\\
10&: A<= 10\; // (E)\\
11&: A<= 11\; // (F).
\end{aligned}
\end{equation*}
Even though signals $C, D, E, F$ have PC scores of zeros by definition (we cannot control any bits of them), we can fully control $A$ if $B$ is under control.
In this case, all four possible 2-bit combinations can be arbitrarily assigned to $A$ through $B$.
Equation~(\ref{eq:casestatement}) helps us cover the assignments involved with constants by assigning to these constants a temporary score when calculating the resulting PC score in conditional assignments.
% However, Equation~(\ref{eq:casestatement}) is not applicable when the select signal $S_B$ is not fully controllable.
% Please note that there is an exception in case F where $S_C$ and $S_D$ are 0 but $S_A$ is 1 as signal $A$ and $B$ have all the values that signal $A$ can possibly have, e.g., 0 and 1.
% However, in an $n$-bit scenario where n is not equal to 1, this function still works.
% For instance, in the statement below:
% \begin{equation*}
%     A <= (B?)\;3'b011\;:\;3'b111,
% \end{equation*}
% where $A$ could only be $011$ or $111$. 
% Even if we can fully control $B$ to select the value for $A$, we still claim the PC score of $A$ is 0 as we cannot actually control the value of $A$, which is decided by designers already.

% \wk{how about a case statement that covers all the scenarios?}
% \wk{exceptions: 0 and 1, talk about an extension to $n$-bit5}

\subsection{Comparison Operator}
Aside from the assignment conditioned on a signal, we often find assignments conditioned on a comparison result, e.g., 
\begin{equation*}
    {A} <= (sig1\; OO\; sig2)\;? \;C:\;D,
\end{equation*}
where $OO$ represents any operator that can fit in this statement, including logical operators and comparison operators, e.g., $||$, $\&\&$, $==$, $>=$.
In this case, we can assume that the comparison result is assigned to signal $B$:
\begin{equation*}
    B = (sig1\; OO\; sig2),
\end{equation*}
where $S_B$ represents the probability that the condition within the parenthesis is true, and a 1-bit signal is used to represent the result.
The resulting PC scores of the logical operators (\textit{$!,\;||,\; \&\&$}) are shown in Table~\ref{tab:operator}. % and scores of the comparison operators are all $\frac{S_1+S_2}{2}$. % shown in Table~\ref{tab:comparison_operator}.
Next, we explain how we derive the score of comparison operators.
We first assume that all these signals are 1-bit each.

\textit{Equal to}: $sig1$ equals $sig2$  can be represented in this way:
% \vspace{2em}
% \newpage
\begin{equation*}
\resizebox{0.91\hsize}{!}{%
    % \begin{aligned}
       $B = (sig1 == sig2) \equiv (\eqnmarkbox[blue]{term1}{(sig1\;\&\&\;sig2)}\;||\; \eqnmarkbox[purple]{term2}{(!sig1\; \&\&\; !sig2)})$
    % \end{aligned}
    }
\end{equation*}
\annotate[yshift=-0.1em]{below}{term1}{term1}
\annotate[yshift=-0.1em]{below}{term2}{term2}

\noindent By combining the defined score functions of the basic operators, we obtain the score $\frac{S_1+S_2}{2}$ for both term 1 and term 2.
By applying the \textbf{OR} operation of these two terms, we have:
\vspace{2.2em}
\begin{equation*}
    \begin{aligned}
        S_B &= \frac{\eqnmarkbox[blue]{score1}{\frac{S_1+S_2}{2}}+\eqnmarkbox[purple]{score2}{\frac{S_1+S_2}{2}}}{2}\\
        % &= \frac{S_1+S_2}{2}
    \end{aligned}
\end{equation*}
\annotate[yshift=1.8em]{above}{score1}{score for term1}
\annotate[yshift=0.7em]{above}{score2}{score for term2}
\begin{equation*}
\begin{aligned}
        &= \frac{S_1+S_2}{2}
        \end{aligned}
\end{equation*}
% Please note that the PC score of logical operator \textit{not ($!$)} is the same as the basic operator \textbf{NOT}.
% \begin{table}[hbtp!]
%     \centering
%      \begin{tabular}{|c|c|}
%     \hline
%         Operator & Resulting PC score \\\hline
%         % $!$ & $S_1$\\ \hline
%         $\&\&$ & $min(S_1,S_2)$\\ \hline
%         $||$ & $max(S_1,S_2)$\\ \hline
%     \end{tabular}
%     \caption{Resulting PC score of logical operators}
%     \label{tab:logic_operator}
% \end{table}

\noindent We use logical expressions to represent these comparisons:\\
% \textit{Not equal to}
% \begin{equation}
%     (sig1\; !=\; sig2) \equiv ((sig1\;\&\&\;!sig2)\;||\;(!sig1\;\&\&\;sig2))
% \end{equation}
\textit{Greater than}
\begin{equation*}
    (sig1 > sig2) \equiv (sig1\;\&\&\;!sig2)
\end{equation*}
\textit{Less than}
\begin{equation*}
    (sig1 < sig2) \equiv (!sig1\;\&\&\;sig2)
\end{equation*}
% \textit{Greater than or equal to}
% \begin{equation}
%     (sig1 \geq sig2) \equiv (sig1\;||\;!sig2)
% \end{equation}
% \textit{Less than or equal to}
% \begin{equation}
%     (sig1 \leq sig2) \equiv (!sig1\;||\;sig2)
% \end{equation}
Other comparisons can be derived in a similar way and we can obtain the expressions for all comparisons as $\frac{S_1+S_2}{2}$. 
% \begingroup
% \renewcommand{\arraystretch}{1.5} 
% \begin{table}[hbtp!]
% \caption{PC score calculation for comparison operators.}
%     \centering
%      \begin{tabular}{|c|c|}
%     \hline
%         Operator & PC score \\\hline
%         % $!$ & $S_1$\\ \hline
%         $==$ & $\frac{S_1+S_2}{2}$\\ 
%         $!=$ & $\frac{S_1+S_2}{2}$\\ 
%         $>$ & $\frac{S_1+S_2}{2}$\\ 
%         $<$ & $\frac{S_1+S_2}{2}$\\ 
%         $\geq$ & $\frac{S_1+S_2}{2}$\\ 
%         $\leq$ & $\frac{S_1+S_2}{2}$\\ \bottomrule
%     \end{tabular}
%     \label{tab:comparison_operator}
%     \vspace{-2em}
% \end{table}
% \begin{table}[]
% \centering
% \caption{PC score calculation for comparison operators.}
% \label{tab:my-table}
% \begin{tabular}{|c|c|c|c|c|c|c|}
% \hline
% Operator & $==$                & $!=$                & $>$                 & $<$                 & $\geq$              & $\leq$              \\ \hline
% PC score & $\frac{S_1+S_2}{2}$ & $\frac{S_1+S_2}{2}$ & $\frac{S_1+S_2}{2}$ & $\frac{S_1+S_2}{2}$ & $\frac{S_1+S_2}{2}$ & $\frac{S_1+S_2}{2}$ \\ \hline
% \end{tabular}
% \end{table}
% \endgroup

% \wk{extend to $n$-bit $\checkmark$ \\}
\subsection{Comparison for multi-bit signals}
Now, we show how to extend the single-bit comparison expressions to multi-bit signals.
Given two $n$-bit signals $sig1$ and $sig2$, we denote the MSB and the LSB of each as $sig1_n,\; sig1_1\;, sig2_n,$ and $sig2_1$, respectively.
A multi-bit signal comparison can be seen as a logical combination of the comparison between each bit.
For instance, the \textit{Equal} comparison can be formulated as:\\
\textit{Equal to}: 
\begin{equation}
\begin{aligned}
    &(sig1_n == sig2_n)\;\&\&\;(sig1_{n-1} == sig2_{n-1})\;\&\&\;\\
    & \cdot \cdot \cdot \&\&\;(sig1_1 == sig2_1)
    \label{eq:multi-comparison}
\end{aligned}
\end{equation}
By definition, each bit of signals $sig1$ and $sig2$ has a PC score $\frac{S_1}{n}$ and $\frac{S_2}{n}$, respectively.
By applying the corresponding functions of each operator, we obtain the score for Equation~(\ref{eq:multi-comparison}) as $\frac{S_1+S_2}{2n}$, which is the single-bit's expressions divided by $n$.

% \textit{Not equal to}: 
% \begin{equation}
%     (sig1_n \oplus sig2_n)\;||\; (sig1_{n-1} \oplus sig2_{n-1})\;||\; \cdot \cdot \cdot \;||\;(sig1_1 \oplus sig2_1)
% \end{equation}

\textit{Greater than}: 
\begin{equation}
\footnotesize
\begin{split}
        (sig1_n > sig2_n)\;||\;( (sig1_n == sig2_n)\;\&\&\;(sig1_{n-1} > sig2_{n-1}))\;||\;\\
        \cdot\cdot\cdot \;||\;(\cdot \cdot \cdot\;\&\&\;(sig1_2 == sig2_2)\;\&\&\;(sig1_1 > sig2_1))\\
\end{split}
\label{eq:greaterthan}
\end{equation}
Since all the single-bit comparisons and basic operators have the same PC score function (i.e., $\frac{S_1+S_2}{2}$) and each bit of both signals has the same score ($\frac{S_1}{n}, \frac{S_2}{n}$), the final score for Equation~(\ref{eq:greaterthan}) is still $\frac{S_1+S_2}{2n}$.
Other comparisons can be derived in a similar manner.

Note that the comparison result is always a single-bit value (True or False).
Therefore, before performing the comparison between two signals, we need to divide the PC score of each signal by their widths, e.g., $\frac{S_1}{n}, \frac{S_2}{n}$. This normalization process ensures that the comparison result is between $0$ and $1$.

When the comparison is done between a signal and a value, the resulting PC score does not change.
% Similarly, we can prove it from single-bit signals.
Consider the situation where we have 1-bit signals A and B. 
We want to see $A == 1$. 
We can take a signal B as the result of the comparison, e.g., $B <= (A == 1)$. 
Because A is a 1-bit signal, this is the same saying that $B <= A$.
For all single-bit comparison operators %(Table~\ref{tab:comparison_operator}), 
it should be clear that $S_B$ is always equal to $S_A$ because $B$ can take either the value of $A$ or $A$ negated.

% \wk{need modifications}
% \begin{subequations}
% \begin{equation}
% \begin{aligned}
%     B &= (A == 1)\\
%     &= (A)
% \end{aligned}
% \end{equation}
% \begin{equation}
% \begin{aligned}
%     B &= (A == 0)\\
%     &= (!A)
% \end{aligned}
% \end{equation}
% \begin{equation}
% \begin{aligned}
%     B &= (A > 0)\\
%     &= (A)
% \end{aligned}
% \end{equation}
% \begin{equation}
% \begin{aligned}
%     B &= (A < 1)\\
%     &= (!A)
% \end{aligned}
% \end{equation}
% \label{eq:comparevalues}
% \end{subequations}

% \noindent As indicated in Equation (\ref{eq:comparevalues}), $S_B$ is always equal to $S_A$ in a signal-to-value comparison.
% \wk{need modifications}
When comparing a multi-bit signal to a fixed value, we can compare each bit of the signal with the corresponding bit of the fixed value, and combine the results to obtain the overall score. For example, to compare a 2-bit signal A:
\begin{equation}
    \begin{aligned}
        (A  == 10) &\Rightarrow (A_2 == 1)\;\&\&\;(A_1 == 0)\\
        &= (A_2)\;\&\&\;(!A_1)\\
        &= \frac{A}{2}
    \end{aligned}
\end{equation}
The PC score remains unaffected by the signal-to-value comparison. 
Note that it is necessary to normalize the result for maintaining consistency and allowing meaningful comparisons between different operations and evaluations.

% \subsection{Reduction operators}
% For reduction operations, we first need to split an $n$-bit signal into $n$ 1-bit signals whose scores are the original score divided by $n$.
% Then, we can apply the basic operation for these $n$ signals to get a final score.
% Since each bit has the same score, following the functions shown in Table~\ref{tab:operator}, the resulting PC score is always the original score divided by $n$.
% For instance, 
% \begin{equation}
%     \begin{aligned}
%         A &= \&B\\
%         &= B_n\;\&B_{n-1}\&...\&\;B_1\\
%         &= \frac{B}{n}
%     \end{aligned}
% \end{equation}
\section{Experimental Evaluation}
% \subsection{Overview}
To evaluate the proposed approach, we applied our formulation to an IP from the Hack@DAC-21 SoC~\cite{dessouky2019hardfails}, based on OpenPiton~\cite{openpiton}.
We assess our proposed method by generating multiple possible patching options, calculating the resulting PC score for each option, analyzing their resource usage, and investigating potentially patchable weaknesses.
% All experiments were performed on a Linux workstation with 40 2.4 GHZ Intel\textsuperscript{\textregistered} Xeon\textsuperscript{\textregistered} CPUs with 754 GB memory. 
We use Verific libraries (academic license) to parse the design files, capture the data/information flow, and prototype a Python-based tool to calculate the resulting patchability scores with user inputs. %Python 3.11

\subsection{Selected Common Weaknesses for Evaluation}
% To examine the effectiveness of our approach, we
We focus on the \textit{reglk\_wrapper}, slightly modifying the code, as presented in Fig.~\ref{fig:code}.
To assess each patching option, we selected several relevant weaknesses from MITRE's common weakness enumerations (CWEs) and proposed example patches, determining the necessary controllable signals to address each CWE.

\textbf{CWE-1262: }Improper Access Control for Register Interface.
Attackers can modify the internal registers due to the improper access control, which are \textit{reglk\_mem} entries in this case study. To patch this, we need to be able to fully control at least one of these signals: \textit{en}, \textit{we}, \textit{address}, \textit{wdata}, \textit{reglk\_ctrl}, or \textit{reglk\_mem}.      

\textbf{CWE-1231: }Improper Prevention of Lock Bit Modification. The lock bit (\textit{reglk\_ctrl}) can be modified so that the access control policy is corrupted. 
The easiest patch is to control \textit{reglk\_ctrl} to mitigate this issue.

\textbf{CWE-1272: }Sensitive Information Uncleared Before Debug/Power State Transition. To avoid sensitive information being read by malicious users during debug mode, we can either clear the sensitive data during debug mode, or control the read transaction. Therefore, one of these signals need to be patched: \textit{reglk\_mem}, \textit{rdata}, \textit{reglk\_ctrl}, \textit{address}, \textit{en}, \textit{rst\_ni}, \textit{jtag\_unlock}, or \textit{rst\_9}.    

\textbf{CWE-276: }Incorrect Default Permissions. For example, \textit{reglk\_mem} should not be assigned 0s during reset. In this case, we have to control \textit{reglk\_mem} for patching.

\begin{figure}[t]
\begin{minted}
[
frame=lines,
xleftmargin=\parindent,
framesep=2mm,
baselinestretch=0.9,
breaklines,
numbersep=\mintednumbersep,
% bgcolor=LightGray,
fontsize=\scriptsize, 
linenos
]
{systemverilog} 
assign reglk_ctrl_o = {reglk_mem[5], reglk_mem[4], 
reglk_mem[3], reglk_mem[2], reglk_mem[1], reglk_mem[0]}; 
assign reglk_ctrl = reglk_ctrl_i;
assign en = en_acct ? acct_ctrl_i: 0; 
always @(posedge clk_i) begin 
  if(rst_ni && ~jtag_unlock && ~rst_9) begin
    for (j=0; j < 6; j=j+1) begin
      reglk_mem[j] <= 'h0;
    end
  end
  else if(en && we) begin
    case(address[7:3])
 0: reglk_mem[0] <= reglk_ctrl[1] ? reglk_mem[0] : wdata;
 1: reglk_mem[1] <= reglk_ctrl[1] ? reglk_mem[1] : wdata; 
 2: reglk_mem[2] <= reglk_ctrl[1] ? reglk_mem[3] : wdata;
 3: reglk_mem[3] <= reglk_ctrl[1] ? reglk_mem[3] : wdata;
 4: reglk_mem[4] <= reglk_ctrl[1] ? reglk_mem[4] : wdata;
 5: reglk_mem[5] <= reglk_ctrl[1] ? reglk_mem[5] : wdata;
    endcase
  end 
end 
always @(*) begin
  rdata = 64'b0; 
  if (en) begin
    case(address[7:3])
      0: rdata = reglk_ctrl[0] ? 1'b0 : reglk_mem[0]; 
      1: rdata = reglk_ctrl[0] ? 1'b0 : reglk_mem[1];
      2: rdata = reglk_ctrl[0] ? 1'b0 : reglk_mem[2];
      3: rdata = reglk_ctrl[0] ? 1'b0 : reglk_mem[3];
      4: rdata = reglk_ctrl[0] ? 1'b0 : reglk_mem[4];
      5: rdata = reglk_ctrl[0] ? 1'b0 : reglk_mem[5];
    endcase
  end  
end 
\end{minted}
\caption{Code snippet of reglk\_wrapper from Hack@DAC-21.}
\label{fig:code}
\end{figure}
\subsection{Case Study}
To understand how different patching options impact the patchability measure, we explore several patching strategies that select different sets of signals for patching in Table~\ref{tab:score}.
\begin{table*}[htbp!]
\centering
\caption{Theoretical patchability quantification with various patching options. Each patching option is presented with \textit{In} and \textit{Out} columns, where each cell represents the number of controllable bits of each signal.}
\label{tab:score}
\resizebox{0.85\textwidth}{!}{%
\begin{tabular}{@{}|l|C{1.5cm}|C{1.1cm}C{1.1cm}|cc|cc|cc|cc|cc|@{}}
\toprule
Signal Name           & All Fully Patchable & Greedy In   & Greedy Out  & V1 In & V1 Out & V2 In  & V2 Out & V3 In  & V3 Out & V4 In  & V4 Out & V5 In & V5 Out \\ \midrule
rst\_ni             & 1                 & 1         & 1         & 0        & 0         & 0         & 0         & 1         & 1         & 0         & 0         & 1          & 1           \\
jtag\_unlock       & 1                 & 1         & 1         & 1        & 1         & 1         & 1         & 1         & 1         & 0         & 0         & 1          & 1           \\
rst\_9              & 1                 & 1         & 1         & 1        & 1         & 1         & 1         & 1         & 1         & 0         & 0         & 1          & 1           \\
we                 & 1                 & 1         & 1         & 0        & 0         & 1         & 1         & 1         & 1         & 0         & 0         & 1          & 1           \\
address            & 64                & 64        & 64        & 0        & 0         & 0         & 0         & 64        & 64        & 0         & 0         & 64         & 64          \\
wdata               & 32                & 32        & 32        & 0        & 0         & 32        & 32        & 32        & 32        & 0         & 0         & 32         & 32          \\
reglk\_ctrl\_i      & 8                 & 8         & 8         & 0        & 0         & 8         & 8         & 8         & 8         & 0         & 0         & 8          & 8           \\
en\_acct            & 1                 & 1         & 1         & 0        & 0         & 0         & 0         & 1         & 1         & 0         & 0         & 1          & 1           \\
acct\_ctrl\_i       & 1                 & 1         & 1         & 1        & 1         & 1         & 1         & 1         & 1         & 0         & 0         & 1          & 1           \\
reglk\_mem{[}0{]}  & 32                & 32        & 32        & 0        & 0         & 0         & 15.8        & 0         & 24        & 32        & 32        & 0          & 24          \\
reglk\_mem{[}1{]} & 32                & 32        & 32        & 0        & 0         & 0         & 15.8        & 0         & 24        & 32        & 32        & 0          & 24          \\
reglk\_mem{[}2{]}  & 32                & 32        & 32        & 0        & 0         & 0         & 15.8        & 0         & 24        & 32        & 32        & 0          & 24          \\
reglk\_mem{[}3{]}  & 32                & 32        & 32        & 0        & 0         & 0         & 15.8        & 0         & 24        & 32        & 32        & 0          & 24          \\
reglk\_mem{[}4{]} & 32                & 32        & 32        & 0        & 0         & 0         & 15.8        & 0         & 24        & 32        & 32        & 0          & 24          \\
reglk\_mem{[}5{]}  & 32                & 32        & 32        & 0        & 0         & 0         & 15.8        & 0         & 24        & 32        & 32        & 0          & 24          \\
en                 & 1                 & 1         & 1         & 0        & 0.5         & 1         & 1         & 0         & 1         & 0         & 0         & 0          & 1           \\
reglk\_ctrl       & 16                & 16        & 16        & 0        & 8         & 0         & 8         & 0         & 8         & 0         & 0         & 0          & 8           \\
rdata             & 32                & 0         & 32        & 0        & 0         & 0         & 11.8        & 0         & 18        & 0         & 8         & 32         & 32          \\
reglk\_ctrl\_o    & 112               & 0         & 112       & 0        & 0         & 0         & 94.5        & 0         & 112       & 0         & 112       & 112        & 112         \\ \midrule
Investment (bits)          & 463               & \multicolumn{2}{c|}{319} & \multicolumn{2}{c|}{3} & \multicolumn{2}{c|}{45}  & \multicolumn{2}{c|}{110} & \multicolumn{2}{c|}{192} & \multicolumn{2}{c|}{254}    \\
Output Score (bits)                  & 463               & \multicolumn{2}{c|}{463} & \multicolumn{2}{c|}{3.5} & \multicolumn{2}{c|}{253.8} & \multicolumn{2}{c|}{393} & \multicolumn{2}{c|}{312} & \multicolumn{2}{c|}{407}    \\
Normalized Score   & 1                 & \multicolumn{2}{c|}{1}   & \multicolumn{2}{c|}{0.2}  & \multicolumn{2}{c|}{0.6}   & \multicolumn{2}{c|}{0.9}   & \multicolumn{2}{c|}{0.4}   & \multicolumn{2}{c|}{0.9} \\
Patchable CWEs   & -                & \multicolumn{2}{c|}{1262, 1231, 1272, 276}   & \multicolumn{2}{c|}{1272}  & \multicolumn{2}{c|}{1262, 1231, 1272}   & \multicolumn{2}{c|}{1262, 1231, 1272}   & \multicolumn{2}{c|}{1262, 1272, 276}   & \multicolumn{2}{c|}{1262, 1231, 1272} \\ \bottomrule
\end{tabular}
}
\end{table*}
% The greedy patching strategy controls all input and internal signals to achieve higher patchability.
% V1--V4 represent different examples of a designer's strategy based on some prior knowledge or other heuristics. % where a designer indicates their intent by selecting control signals %from other IPs 
% or signals containing security-relevant information as patchable.
% For patching infrastructures that only patch at the interconnect level, V5 is designed to control all the interface signals.
% ``\textit{All Fully Patchable}'' option selects all signals to be fully patchable, so the score of each signal is the same as their signal widths.
Each patching option has \textit{In} and \textit{Out} columns. 
\textit{In} shows the signals designers want to patch, and \textit{Out} shows the resulting PC score after calculation.
As we do not support partial control, i.e., controlling a subset of a signal, scores in \textit{In} column are either 0 or equal to the signal widths (i.e., the corresponding value in \textit{All Fully Patchable} column).
Investment is the sum of the \textit{In} scores, indicating the number of bits that are directly controlled by the patching design, and also determines the resource investment of a patching block.
While the patching block's building cost does not scale linearly with the number of bits it can patch, a higher investment leads to a larger patching block, resulting in increased resource use.

Output score is the sum of the scores in \textit{Out} column, representing how many bits are controllable in a design. 
We divide the score in each cell of the \textit{Out} column by the width of the corresponding signal, then sum up the resulting values for all signals, and finally divide the sum by the total number of signals to produce a normalized score. 
This score represents the average controllability per signal and accounts for %can help to eliminate differences caused by 
different signal widths when we compare different designs.

\subsection{Comparing Different Patchable Options}
\label{sec:analysis}
% In this section, we analyze the effectiveness of different patching strategies using our proposed patchability formulation. 
% Without knowing what signals deserve patching, t
The greedy option makes all signals patchable except for the output, resulting in a high patchability and the %output and normalized scores, but it also incurs the 
highest resource investment. 
V1 selects some control signals that may appear helpful for patching, but they do not produce much PC to the downstream signals and are insufficient to patch many CWEs. 
V2 and V3 have more signals chosen for patchability, achieving normalized scores of 0.6 and 0.9, respectively. 
These options cost less than one-fourth of the total bits (463) and can patch up to three CWEs. 
Without our patchability guidance, designers might have chosen  V4 to protect sensitive information by patching the \textit{reglk\_mem} signal. 
However, compared to  V3, V4 costs more and achieves lower patchability scores. 
V5 is designed for scenarios where patching blocks can only be integrated at the interconnect level, making all signals at the interface patchable. 
V5 achieves similar scores to V3 but requires more bits in investment and cannot patch more CWEs than V3, demonstrating the usefulness of improving internal patchability. 

Fig.~\ref{fig:code} shows that when signals in the sensitivity list of the first always block, i.e., \textit{rst\_ni}, \textit{jtag\_unlock}, \textit{rst\_9}, \textit{en}, \textit{we}, \textit{reglk\_ctrl}, \textit{address}, and \textit{wdata} become fully controllable, arbitrary values can be assigned to \textit{reglk\_mem}  through write transactions.
Therefore, a high PC score of 24 is assigned to \textit{reglk\_mem} under V3's configuration.
By using the proposed formulation, we can precisely capture the relationship between signals and propagate PC scores from inputs to outputs.
Previously, balancing resource usage spent on patching architectures for each target IP was challenging, given no way to quantify the resulting patchability. 
Our approach quantifies patchability by analyzing the structure of the design and how the data flows, making it easier to balance the resource usage while improving patchability for each block.
As a result, V3 achieves a normalized score of 0.9 while utilizing 65\% fewer resources compared to the greedy approach.

% \begin{enumerate}
%     \item show an example of using this formulation
%     \item assign signals security values to be combined with the formulation such that the selected signals not only have higher controllability but also security meaning\\
%     how many bits controlled vs. the summation of the \ac{PC} scores of the whole design
%     \item how long it takes to run 
%     \item compare a selection of patching strategies (different combinations) to showcase how many more CWEs we can patch?
% \end{enumerate}

\section{Discussion\label{sec:discussion}}
% Please note that w
We assumed that a given CWE is considered patchable if at least one of the signals that we identified as required to be patchable is ``fully''  controllable. 
% This means that even if a patching option has partially controllable signals, it is still considered not fully patchable for that specific CWE.
For example, even though V3 can control 24 out of 32 bits of \textit{reglk\_mem}, CWE-276 is still not considered patchable.
However, in practice, small modifications to signals may be sufficient to prevent an attack, even if they are not fully controllable. 
For instance, in Fig.~\ref{fig:code}, only two bits of \textit{reglk\_ctrl} are used to lock registers, and these bits can be fully controlled by controlling the input signal \textit{reglk\_ctrl\_i}.
In this scenario, these two bits have different security meanings compared to other bits.
Therefore, even though \textit{reglk\_ctrl} is only partially controllable, it can still prevent an exploit of CWE-276. % attack.

To further refine the patchability quantification formulation, it is essential to separate the controllability of signals to ``0'' and ``1''.
For instance, in Fig.\ref{fig:code}, to prevent CWE-1262 where \textit{reglk\_ctrl} is set incorrectly so that attackers can modify \textit{reglk\_mem} through write transactions, one possible patch is to control \textit{reglk\_ctrl} to mitigate this issue.
In this case, it is only necessary to be able to set \textit{reglk\_ctrl} to 1 to block write transactions, and controlling it to 0 is unnecessary.
Our future work will model the controllability of signals in a more granular way, separately quantifying the ability to control a signal to 0 or 1 as needed for a patch. 
This will allow us to quantify patchability more precisely and develop more efficient patching strategies.

% However, in this study, as we do not consider partial controllability, CWE-276 is classified as not patchable for  V2, V3, and V5 because our score does not immediately indicate if the critical bits are under control. % we cannot tell if the controlling bits are under control. % or not by the formulation.
% Also, as individual bits within a signal array can hold distinct security implications, patching different bits can result in varying patchability levels. 

% In addition to the aforementioned scenarios, partially controlling signals can also be effective when only a specific range of values is required for patching.
% For instance, consider the case statement on line 28 in Fig.\ref{fig:code} where \textit{address} determines which case item to execute.
% In this case, setting \textit{address} to a value greater than 5 can force \textit{rdata} to be 0 if more than 2 bits within \textit{address[7:3]} are controllable, i.e., setting all controllable bits to 1 can guarantee the value greater than 5.
% Considering partially controlling signals is crucial in making patchability quantification more practical and efficient. 
% By doing so, redundant bits can be saved, and designers can select specific bits for patching in specific scenarios, leading to varying patchability levels.

Note that the choice of example CWEs in this study is independent of the patching options; in practice, each designer will anticipate potential issues in an SoC on a case-by-case basis. % as it is impossible to anticipate the kind of issues that can occur in a System-on-Chip before identifying a security issue,.
While we reported the number of patchable CWEs as an indicator of a patching option's effectiveness, a high PC score does not guarantee coverage over all possible bugs.  % it is not enough to fully represent its patchability. 
% Likewise, higher theoretically 
% a high(er) PC score does not guarantee the ability to patch more bugs in the field. 
Still, we can interpret a relatively higher score when comparing different options as characterizing a design with a higher probability of successful patching. 
Additionally, other than using each signal's width to represent the score, an alternative approach is to assign a security value to each signal that reflects its relative importance. 
% By incorporating these security values into the PC score, we can better quantify the security level of the patching design. 
This method is particularly useful when system integrators have a deep understanding of the target design and can assign appropriate security values to each signal.

While we focused on PC for brevity, PO is important in patching to capture how well a design can be monitored for misbehavior. 
Both PC and PO are essential components in accurately assessing a system's patchability and developing effective patching strategies.
Even though high PC and PO empower patching designs to change a design's behavior, the security of the patching design itself is a critical consideration. 
Therefore, ensuring the secure delivery of patches is essential to prevent attackers from intercepting and tampering with the patch before it is received. 
% While a full discussion is beyond the scope of this paper, we note that there are several measures that can be taken to secure the patch delivery process, including the use of secure communication channels~\cite{liyanage2017enhancing}, as well as digital signatures and checksums~\cite{meylan2020study} to verify authenticity and integrity. 
Additionally, it is important to restrict patch creation to verified IP vendors or system integrators to prevent unauthorized access or modification. 

\section{Related Prior Work~\label{sec:related}}
Several approaches have been proposed to incorporate hardware-based patching mechanisms into SoCs for in-field IP patching. 
Nath et al.~\cite{nath2018system} introduced a centralized hub utilizing design-for-debug infrastructure/wrappers to monitor transactions and verify security policies at run-time.
They also presented a security architecture for SoCs that involve IP communication through a Network-on-Chip (NoC) fabric~\cite{nath2020resilient}. 
Tan et al.~\cite{tan2020toward} proposed decentralized patching blocks that can be reprogrammed and integrated into the IP interface. 
Each block stores patch programs that monitor IP behavior and override signals at the interface if buggy behavior is detected. 
Liu et al.~\cite{liu2022hardware} also proposed a decentralized patching logic and developed a formulation to help system integrators engineer the entire patching infrastructure at the system level. 
% \wk{
While prior works focus on the design of system-level patching architectures for IP design bugs and security policies, our work provides a systematic evaluation of "patchability" at the IP level. 
This supports system integrators in engineering the patching architecture with finer-grained assessment.
% }
The work by Guo et al.~\cite{guo2019qif} quantifying information flow in hardware systems inspired us in quantifying patchability at the RTL.
\section{Conclusion\label{sec:conclusion}}
% This work introduced an approach to quantify an IP block's patchability, a metric that enables designers to compare patchable hardware options based on patching controllability (PC) and patching observability (PO). 
% We presented a novel formulation that quantifies patchability using probabilistic and heuristic methods to estimate the number of controllable and observable bits in a design.
% Our approach enables system integrators and designers to assess patching blocks, optimize patching configurations for each IP, and strike a balance between patchability and resource investment. 
% By capturing the dataflow relationships between signals, our formulation provides a useful tool for evaluating patchability at the RTL level, improving the security and reliability of hardware systems.

This paper proposed a new approach for measuring the patchability of IP blocks. This metric helps designers compare different hardware options based on their patching controllability and observability. 
Our method uses probabilistic and heuristic methods to estimate controllable and observable bits in a design, considering dataflow relationships between signals. 
This tool can optimize patching configurations at the RTL for each IP while improving the security and reliability of hardware systems.

% \section*{Acknowledgments}
% The authors would like to thank Industry Liaisons Peilin Song, Amitabh Das, and Sohrab Aftabjahani, for their support and insightful discussions on this work. Co-author Jason M. Fung is also an Industry Liaison. 
% The views presented in the paper are the authors' and do not necessarily represent the views of Intel Corporation nor its subsidiaries. 
% \section*{Acknowledgment}
% The views presented in the paper are the authors' and do not necessarily represent the views of Intel Corporation or its subsidiaries. 

% \clearpage

\IEEEtriggercmd{\balance}
\IEEEtriggeratref{20}

\bibliographystyle{IEEEtran}
\bibliography{ref}
\end{document}